\documentclass[lettersize,journal]{IEEEtran}

\usepackage{blindtext}
\usepackage[utf8]{inputenc}
\usepackage[english]{babel}
\usepackage{booktabs}
\usepackage{multirow}
\usepackage{amssymb}
\usepackage[noend]{algorithmic}
\usepackage{algorithm}
\usepackage{cite}
\usepackage{array}
\usepackage{textcomp}
\usepackage{stfloats}
\usepackage[final]{graphicx}
\usepackage{xcolor}
\usepackage{amsmath}
\usepackage{mathtools}
\usepackage{comment}
\usepackage{color,soul}

\DeclareMathOperator{\EX}{\mathbb{E}}
\usepackage{stackengine}

\usepackage{booktabs}
\usepackage{caption}
\usepackage[cmintegrals]{newtxmath}
\usepackage{balance}
\usepackage[export]{adjustbox}
\usepackage[caption=false,font=normalsize,labelfont=sf,textfont=sf]{subfig}

\usepackage{siunitx}
\makeatletter
\providecommand\add@text{}
\newcommand\tagaddtext[1]{%
  \gdef\add@text{#1\gdef\add@text{}}}%
\renewcommand\tagform@[1]{%
  \maketag@@@{\llap{\add@text\quad}(\ignorespaces#1\unskip\@@italiccorr)}%
}
\makeatother

\newcommand{\mybibliography}{\bibliography{jour_short,conf_short,references.bib}}

\begin{document}
\bstctlcite{IEEEexample:BSTcontrol}
\title{Energy-Aware Federated Learning with Distributed User Sampling and Multichannel ALOHA}%

\author{Rafael~Valente~da~Silva,
Onel~L.~Alcaraz~López,
and~Richard~Demo~Souza
\thanks{This research has been supported by the Finnish Foundation for Technology Promotion, Academy of Finland (6G Flagship program under Grant 346208), the European Commission through the Horizon Europe/JU SNS project Hexa-X-II (Grant Agreement no. 101095759), CNPq (402378/2021-0, 401730/2022-0, 305021/2021-4) and RNP/MCTIC 6G Mobile Communications Systems (01245.010604/2020-14).}
\thanks{R. V. Silva and Onel L. Alcaraz López are with the Centre for Wireless Communications (CWC), University of Oulu, 90570 Oulu, Finland \{rafael.valentedasilva@oulu.fi, onel.alcarazlopez@oulu.fi\}.}%
\thanks{R. D. Souza is with the Department of Electrical and Electronics Engineering of the Federal University of Santa Catarina, Florianópolis, Brazil \{richard.demo@ufsc.br\}.}}%

\maketitle

\begin{abstract} 
Distributed learning on edge devices has attracted increased attention with the advent of federated learning (FL). Notably, edge devices often have limited battery and heterogeneous energy availability, while multiple rounds are required in FL for convergence, intensifying the need for energy efficiency. Energy depletion may hinder the training process and the efficient utilization of the trained model. To solve these problems, this letter considers the integration of energy harvesting (EH) devices into a FL network with multi-channel ALOHA, while proposing a method to ensure both low energy outage probability and successful execution of future tasks. Numerical results demonstrate the effectiveness of this method, particularly in critical setups where the average energy income fails to cover the iteration cost. The method outperforms a norm based solution in terms of convergence time and battery level. 
\end{abstract}


\section{Introduction}
\label{sec:intro}

\IEEEPARstart{F}{ederated} learning (FL) has emerged as a prominent research topic within the wireless communication community, gaining significant attention in recent years~\cite{Wahab2021}. In FL, edge devices collaboratively train a global model by only sharing local model updates, which provides a higher protection against the exposure of sensitive data, such as surveillance camera images, geolocation data, and health information. However, such collaborative training requires multiple communication rounds, raising spectral and energy efficiency concerns~\cite{Wahab2021}. The latter is particularly important for edge devices, given their inherent energy limitations. 

The sixth generation ($6$G) of wireless systems targets \mbox{$10$-$100$} times more energy efficiency than $5$G, which is critical for supporting massive Internet of Things (IoT) networks~\cite{Guo2021}. Such demanding vision requires a meticulous design of the communication system, where medium access control (MAC) mechanisms play a major role. Grant-free random access protocols, such as slotted ALOHA (SA) with multiple channels, are suitable candidates for massive IoT applications, since control signaling is much reduced. Moreover, energy availability must be considered to support self-sustainable networks, in which   \textit{energy neutrality} \cite{Albano2021}, balancing availability and expenditure of energy resources, is essential.

Existing literature on FL indirectly addresses spectral and energy efficiency by optimizing the convergence time, leveraging informative updates from users~\cite{Choi2020, daSilva2022} or the relationship between local and global models~\cite{Wu2022}, reducing the required number of iterations. These approaches often overlook the initial battery levels of different devices, which can result in energy depletion during the training process and hinder the overall progress. Even if the training process is not impeded, the remaining energy may be insufficient for the execution of future tasks and the utilization of the trained model.

This letter considers the use of EH devices, which eliminate the need for frequent battery replacement \cite{López2021}, while also allow energy neutrality. Prior works in \cite{Zeng2022, Hamdi2022} considered some sort of energy income for FL networks. In \cite{Zeng2022}, a wireless-powered FL system is considered and the tradeoff between model convergence and the transmission power of the access point is derived. The authors in \cite{Hamdi2022} consider EH devices with multiple base stations (BS) and propose a user selection algorithm to minimize the training loss. However, \cite{Zeng2022, Hamdi2022} overlook the residual energy in the devices at the end of the training process and the energy imbalances among users, which are considered in this letter. Moreover, they do not consider a random access protocol and massive IoT settings. We present a novel energy-aware user sampling technique for a FL network under a multichannel SA protocol. The proposed method enables users to make informed decisions regarding their participation in an iteration, controlling the computation cost. Numerical results corroborate the effectiveness of our method. In critical energy income setups, lower error and higher energy availability can be achieved compared to \cite{Choi2020}, which solely considers the informativeness of updates. We can achieve an error $46.72$\% smaller, while maintaining $37$\% more energy in a network of $100$ devices, while the the performance gap increases with the number of deployed devices.
\section{System Model}
\label{sec:model}
\begin{figure}[] 
    \centering
    \includegraphics[width=0.5\textwidth, trim=0.1mm 0.2mm 0.4mm 0.5mm,clip]{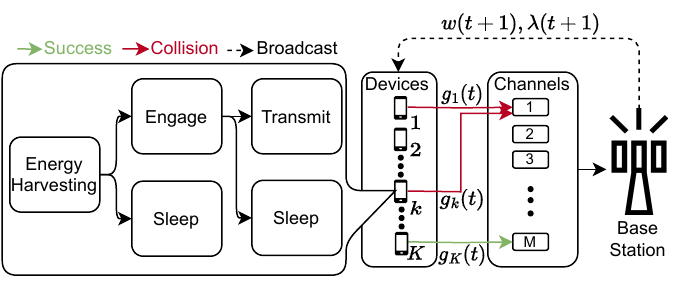} 
    \caption{Users begin the iteration by harvesting energy. Then, a user may engage by computing its local model update $\mathbf{g}_k(t)$. A user can either transmit or withhold its update. Transmissions occur through one of $M$ channels using SA. If more than one user access the same channel, there is a collision.}
    \label{fig:system_model}
\end{figure}
Consider a wireless network comprising $K$ users, indexed as $k \in \mathcal{K} = \{1, 2, \ldots, K\}$, a BS, and $M$ orthogonal channels. Each user has a dataset $\mathcal{D}_{k}=\{\mathbf{x}_{k}, \mathbf{y}_{k}\}$ associated with its respective local model. Here, $\mathbf{x}_k$ is the unlabeled sample vector, with size $L \times 1$, and $\mathbf{y}_{k}$ is the ground truth vector for supervised learning. The common goal of every device is to minimize a global loss function $F(\mathbf{w})$ as
\begin{equation}
     \min_{\mathbf{w}} \dfrac{1}{K}\sum\limits_{k=1}^K f_k(\mathbf{w}),
     \label{eq:problem_global_loss_function}
\end{equation}
where $f_k(\mathbf{w}) = \ell(\mathbf{w}, \mathbf{x}_{k}, \mathbf{y}_{k})$ is the local loss function for the $k$-th user and $\textbf{w}$ is the global model. In FL, the problem in \eqref{eq:problem_global_loss_function} is tackled by distributively minimizing $f_k(\mathbf{w})$ over iterations, which yields a local model update $\mathbf{g}_k(t) = \nabla f_k(\mathbf{w}(t))$ for the stochastic gradient descendent method. To ensure collaborative learning, each user transmits $\mathbf{g}_k(t)$ to the BS, which employs an aggregation function to update the global model. Here, we consider FedAvg \cite{Konecny2017}, thus, the global model is updated as
\begin{equation}
     \mathbf{w}(t + 1) = \mathbf{w}(t) - \mu\sum\limits_{k \in \mathcal{K}}d_k \mathbf{g}_k(t),
     \label{eq:global_model_updates}
\end{equation}
where $\mu > 0$ is the learning rate and $d_k = |\mathcal{D}_k|/\sum_{k^{'}=1}^K |\mathcal{D}_{k^{'}}|$. Then, the BS broadcasts $\mathbf{w}(t+1)$ for all users.

From \eqref{eq:global_model_updates}, we can observe that the size of the learning step is directly affected by the norm of the local update $||\mathbf{g_k}(t)||$, which quantifies how informative the update is. In \cite{Choi2020}, the authors present a method to adaptly decide the transmission probability of users based on the local update norm given by
\begin{equation}
    p_{\text{tx}, k}(t) = \max (\min (e\ln{||\mathbf{g_k}(t)||} - \lambda(t), 1), 0).
    \label{eq:p_tx}
\end{equation}
In this context, $\lambda(t)$ serves as a feedback signal that ensures an efficient utilization of the $M$ orthogonal channels in a multichannel SA setup \footnote{As discussed in~\cite{Yao2023}, transmission errors (or collisions) may compromise the FL performance. However, following~\cite{Choi2020}, the considered network maximizes the utilization of the available resources.}. The value of $\lambda(t)$ is determined by
\begin{equation}
    \lambda(t)  = \lambda(t - 1) + \mu_1(\hat K - M),
    \label{eq:feedback}
\end{equation}
where $\mu_1$ is a step size and $\hat K \le K$ is the number of transmissions that occurred at the previous iteration. 

Note that this method does not consider the, potentially limited, energy availability at the devices. For instance, an EH user could repeatedly transmit and drain its battery in the process, rendering the execution of future tasks impossible. To mitigate this, we introduce a sleep probability and consider an strategy depicted in Fig.~\ref{fig:system_model} and based on the following steps.
\begin{enumerate}
    \item \textbf{Energy Harvesting:} At the start of an iteration, each device harvests $\zeta_k(t)$ Joules of energy and stores in the battery if its capacity allows, being $\zeta_k(t)$ a random variable with a predefined distribution. 
    \item \textbf{Engagement:} Each user decides whether to engage in the iteration with a sleep probability 
    \begin{align}
        p_{\text{s}, k}(t) = 1 - \alpha\tfrac{B_k(t)}{B_{\text{max}}},
        \label{eq:sleep_probability}
    \end{align}
    where $\alpha$  is a constant, $B_k(t)$ is the current battery level, and $B_{\text{max}}$ is the battery capacity, which is the same for all devices. We propose this sleep probability to equalize the battery charge of all devices over time. The awaken users receive the global model $\mathbf{w}(t)$ from the BS and compute their local model updates $\mathbf{g}_k(t)$.
    \item \textbf{Informative Multi-Channel SA:} Users transmit $\mathbf{g}_k(t)$ with a probability given by \eqref{eq:p_tx}. Transmissions occur through a randomly chosen channel among $M$ channels. A transmission is only successful if there is no collision.
    \item \textbf{Global Model Updates}: Following \eqref{eq:global_model_updates} the BS aggregates the local updates and broadcasts $\mathbf{w}(t+1)$ and $\lambda(t+1)$, which are assumed to be collision-free.
\end{enumerate}

Following this procedure, the battery evolution model is
\begin{align}
    B_k(t) &= B_k(t-1) + \min(\zeta_k(t), B_{\text{max}} - B_k(t-1)) \nonumber \\
    &- \delta_{\text{e}, k}(t)(E^\text{cmp}_k + E^{\text{rx}}_k) - \delta_{\text{tx}, k}(t) E^{\text{tx}}_k,
    \label{eq:battery_evolution_model}
\end{align}
where $\delta_{\text{e}, k}(t)$ and $\delta_{\text{tx}, k}(t)$ are indicator functions representing user engagement and transmission, respectively. They are equal to 1 when the corresponding event occurs and 0 otherwise. Additionally, $E^\text{cmp}_k$, $E^{\text{rx}}_k$, and $E^{\text{tx}}_k$ are the computation, reception, and transmission energy costs, respectively, whose models are presented in Section \ref{sec:energy_models}. Moreover, it is crucial to choose a precise value for $\alpha$ in \mbox{step 2)} to ensure the proper functioning of the network, which is discussed in Section \ref{sec:proposed_method}.

\section{Energy Consumption Models} \label{sec:energy_models}
\subsection{Local-Computation Model}

The computation complexity of a machine learning algorithm can be measured by the number of required floating point operations (FLOPs). Let $W$ denote the number of FLOPs per data sample for a given  model. The total number of FLOPs for the $k$-th user to perform one local update is
\begin{equation}
 G_k = W|\mathcal{D}_k|.
\label{eq:Flops_update}
\end{equation}
Let $f_{\text{clk}, k}$ be the processor clock frequency (in cycles/s) of the $k$-th user and $C_k$ be the number of FLOPs it processes within one cycle. Then, the time required for one local update is
\begin{equation}
    t_k = \dfrac{G_k}{C_kf_{\text{clk}, k}}, \quad \forall k \in \mathcal{K}.
    \label{eq:time_localupdate}
\end{equation}
Moreover, for a CMOS circuit, the central processing unit (CPU) power is often modeled by its most predominant part: the dynamic power \cite{Zhang2013}, which is proportional to the square of the supply voltage and to the operating clock frequency. Moreover, for a low voltage supply, as in our case, the frequency scales approximately linear with the voltage \cite{Zhang2013}. Therefore, the CPU power consumption can be written as \cite{Zeng2022}
\begin{equation}
    P^\text{cmp}_k = \psi_k f_{\text{clk}, k}^3 \quad \forall k \in \mathcal{K},
    \label{eq:power_CMOS}
\end{equation}
where $\psi$ is the effective capacitance and depends on the chip architecture. Based on \eqref{eq:time_localupdate} and \eqref{eq:power_CMOS}, the energy consumption of the computation phase for the $k$-th user is given by
\begin{equation}
    E^\text{cmp}_k = t_k P^\text{cmp}_k = \psi_k \dfrac{G_k}{C_k}f_{\text{clk}, k}^2.
    \label{eq:Energy_cmp}
\end{equation}

\subsection{Transceiver Model}
The energy consumed by the edge devices' transceivers is 
\begin{equation}
    E^{\text{comms}}_k = E^{\text{tx}}_k + E^{\text{rx}}_k + E^{\text{sleep}}_k,
    \label{eq:energy_comms}
\end{equation}
where $ E^{\text{tx}}_k$ ($E^{\text{rx}}_k$) is the energy required to transmit (receive) a local (global) update while $E^{\text{sleep}}_k$ is the consumed energy during the inactive time. Since $E^{\text{sleep}}_k$ is much smaller than $E^{\text{tx}}_k$ and $E^{\text{rx}}_k$, we neglect its impact in the following. 

Considering the transmission of local updates with a radiated power $P_k^{\text{tx}}$, the power consumed by the edge transceivers is can be modeled as \cite{Scaciota2022}
\begin{equation}
    P^{\text{total}}_k = \dfrac{P^{\text{tx}}_k}{\eta} + P_{\text{circ}},
    \label{eq:ptx}
\end{equation}
where $\eta$ is the drain efficiency of the power amplifier (PA), and $P_{\text{circ}}$ is a fixed power consumption that comprises all other transceiver circuits except the PA. Then, the energy required to transmit a local update is
\begin{equation}
    E^{\text{tx}}_k = \dfrac{P^{\text{total}}_k}{R^{\text{tx}}_b}N_k,
\end{equation}
where $N_k$ is the local update size in bits, and $R^{\text{tx}}_b$ is the bit rate in the uplink. Meanwhile, the energy consumed when receiving the global updates is modeled by
\begin{equation}
    E^{\text{rx}}_k = \dfrac{P^{\text{rx}}_k}{R^{\text{rx}}_b}N,
\end{equation}
where $N$ is the global update size in bits, $R^{\text{rx}}_b$ is the bit rate in the downlink, and $P^{\text{rx}}_k$ is the receive power consumption, which includes $P_\text{circ}$. Thus, $P^{\text{rx}}_k$ is slightly greater than $P_\text{circ}$, but usually smaller than $P_k^{\text{total}}$.

\section{Sleep Probability Tuning}\label{sec:proposed_method}
To ensure that a device saves enough energy for future tasks while still participating in the model training, we propose a precise selection of parameter $\alpha$ based on the EH process and the desired battery level at the end of the training. Notice that the expected battery level with respect to $k$ and assuming equal costs for all devices can be obtained from \eqref{eq:battery_evolution_model} as
\begin{align}
\mathbb{E} [B_k(t)] &= \mathbb{E}[B_k(t-1)] + \mathbb{E}[\min(\zeta_k(t), B_{\text{max}} - B_k(t-1))] \nonumber \\
&- \mathbb{E} [\delta_{\text{e}, k}(t)](E^\text{cmp} + E^{\text{rx}}) - \mathbb{E}[\delta_{\text{tx}, k}(t)] E^{\text{tx}} \nonumber \\
&=\EX[B_k(t-1)] + \EX[\min(\zeta_k(t), B_{\text{max}} - B_k(t-1))] \nonumber \\
&- \alpha \dfrac{\EX [B_k(t)]}{B_{\text{max}}}(E^\text{cmp} + E^{\text{rx}}) - p_{\text{tx}, k}(t) E^{\text{tx}},
\label{eq:battery_evolution_expectation}
\end{align}
where $\EX [\delta_{\text{e}, k}(t)] = 1 - p_{\text{s}, k}(t)$ and $\EX [\delta_{\text{tx}, k}(t)] = p_{\text{tx}, k}(t)$. We also consider the expectation of the battery level in $p_{\text{s}, k}$, since we aim to stabilize the average battery level to a fixed threshold $\xi>0$ over time. Therefore, as $t$ tends to infinity, $\mathbb{E} [B_k(t)]$ converges to $\xi$. Using this in \eqref{eq:battery_evolution_expectation} leads to
\begin{align}
    \alpha = \left(E_h - p_{\text{tx}, k}(t) E^{\text{tx}}\right) \dfrac{B_{\text{max}}}{\xi(E^\text{cmp} + E^{\text{rx}})},
     \label{eq:alpha_equal}
\end{align}
where $E_h = \EX[\min(\zeta_k(t), B_{\text{max}} - B_k(t-1))]$ is the average harvested energy. 
Note that the proposed solution requires knowledge of $\zeta_k(t)$ and $B_k(t-1)$ distributions. Although it is reasonable to assume that a device has such knowledge, mathematical tractability of the battery level is challenging. Since the required battery knowledge pertains to a previous time than the energy income, the distributions of these two variables are independent. This allows us to rearrange the expectations and state the average harvested energy as
\begin{align}
    E_h &= \EX[\min(\zeta_k(t), B_{\text{max}} - B_k(t-1))] \nonumber \\
    &= \EX_{\zeta}[\EX_{B}[\min(\zeta_k(t), B_{\text{max}} - B_k(t-1))]]  \nonumber \\
    &\mathrel{\overset{\makebox[0pt]{\mbox{\normalfont\tiny\sffamily (a)}}}{\ge}} \EX_{\zeta}[\min(\zeta_k(t), B_{\text{max}} - \EX[B_k(t-1)])] \nonumber \\
    &\mathrel{\overset{\makebox[0pt]{\mbox{\normalfont\tiny\sffamily (b)}}}{=}} \EX[\min(\zeta_k(t), B_{\text{max}} - \xi)] \nonumber \\
    &= \EX[\zeta_k(t)\, |\, \zeta_k(t) \le B_{\text{max}} - \xi]\text{Pr}\{ \zeta_k(t) \le B_{\text{max}} - \xi\} \nonumber \\
    &+ (B_{\text{max}} - \xi)\text{Pr}\{ \zeta_k(t) > B_{\text{max}} - \xi\}.
    \label{eq:eh}
\end{align}
Since the minimum function is convex, we employed Jensen's inequality in step (a) and from step (b) onward we consider $t \to \infty$, thus $\EX [B_k(t - 1)] = \xi$.

Since $p_{\text{tx}, k}(t)$ is not known a priori, and to allow deviations of the energy stored in the battery about $\xi$, we use $\EX [p_{\text{tx}, k}(t)]$ in \eqref{eq:alpha_equal} instead of $p_{\text{tx}, k}(t)$. According to \eqref{eq:feedback}, out of the $K$ users, $M$ updates per iteration are transmitted on average to the BS, thus, $\EX [p_{\text{tx}, k}(t)] = M/K$. Then, with \eqref{eq:eh} and \eqref{eq:alpha_equal} we have
\begin{align}
    \alpha &\ge \left(\EX_k[\min(\zeta_k(t), B_{\text{max}} - \xi)] - \dfrac{M}{K} E^{\text{tx}}\right) \dfrac{B_{\text{max}}}{\xi(E^\text{cmp} + E^{\text{rx}})}.
    \label{eq:alpha_inequality}
\end{align}

At the beginning of the training process, the BS broadcasts the value of $\alpha$ solved by assuming equality in~\eqref{eq:alpha_inequality}. 

\subsection{Mean EH Knowledge}

We also consider a simpler variation of the method where we exploit only the average EH information, i.e., we use \mbox{$E_h = \EX[\zeta_k(t)]$} and \mbox{$\EX [p_{\text{tx}, k}(t)] = M/K$} in \eqref{eq:alpha_equal}, thus
\begin{align}
    \alpha = \left(\EX[\zeta_k(t)] - \dfrac{M}{K} E^{\text{tx}}\right) \dfrac{B_{\text{max}}}{\xi(E^\text{cmp} + E^{\text{rx}})}.
     \label{eq:mean_knowledge}
\end{align}
The energy mean knowledge (EMK) approach in \eqref{eq:mean_knowledge} disregards the impact of the maximum battery capacity, different from the energy distribution knowledge (EDK) in \eqref{eq:alpha_inequality}.

\section{Simulation Results}\label{sec:results}
We analyze the performance of the proposed method compared to the Largest Updates' Norms (LUN) baseline, where users transmit the updates with the largest norms according to \cite{Choi2020}. Additionally, to illustrate the necessity of the adaptive control presented in \eqref{eq:p_tx} and \eqref{eq:feedback}, we include a baseline method that assigns a uniform transmission probability $p_{\text{tx}, k} = M/K$ to all users (to distinguish, we use the acronym AC for adaptive control). We assume a linear regression problem with the following loss function: $f_k(\mathbf{w}) = 0.5 | \mathbf {x}_{k}^ {\mathrm {T}} \mathbf {w}(t) - {y}_{k}|^{2}$ \cite{Choi2020}, where $\mathbf {x}_{k}\thicksim\mathcal{N}(\mathbf{v}_k, \mathbf{I})$, \mbox{${y}_{k} =  \mathbf {x}_{k}^ {\mathrm {T}} \mathbf {w}$}, and $\mathbf {w}\thicksim\mathcal{N}(\mathbf{0}, \mathbf{I})$. Note that $\mathbf {w}(t)$ are the training weights, while $\mathbf {w}$ corresponds to the true weights. Also, parameter $\mathbf{v}_k\thicksim\mathcal{N}(0, \mathbf{\beta}_k)$ is utilized to generate a non-IID dataset, with $\mathbf{\beta}_k = \mathbf{I}$ indicating the \mbox{non-IID} degree.

Similar to \cite{Hamdi2022}, the energy income at each user is modeled by a compound Poisson stochastic process, i.e., the interarrival time is modeled by an exponential distribution with rate $r$ and the amount of energy harvested in each arrival is modeled by a Poisson process with parameter $m/r$, thus, $\EX_{t}[\zeta_k(t)] = m$. This model is defined by discrete units of energy. We scale one unit of energy to the total cost of an iteration in J, i.e., $E^{\text{comms}}_k  + E^\text{cmp}_k$. Unless stated otherwise, we set $r = 0.02$ and $m = 0.2$ units of energy. Note that $r$ is the mean of the exponential distribution, corresponding to an energy arrival every $50$ iterations on average, similar to~\cite{Aygün2022}. Moreover, we set $K = 100$, $M = 10$, $L = 10$, $\mu = 0.01$, and $\mu_1 = 0.1$ as in~\cite{Choi2020}, while $P^{\text{tx}}_k = 3.3$~dB, $P^{\text{rx}}_k = 1.9$~mW, $\eta = 0.33$, $P_{\text{circ}} = 1.33$~mW, which correspond to a BLE transceiver~\cite{Tamura2020}. Moreover, \mbox{$R^{\text{tx}}_b = R^{\text{rx}}_b = 1$~Mbps}, $W = 4L$, $f_{\text{clk},k} = 0.25$~GHz, $C_k = 20$~\cite{Xu2021}, and the effective capacitance is $\psi_k = 10^{-20}$~\cite{He2022}, while the initial battery level of the devices is given by a uniform distribution $U(0, B_{\text{max}})$, where $B_{\text{max}} = 10^{-1}$~J.

\begin{figure}[t!] 
    \centering
    \captionsetup[subfigure]{oneside,margin={0.8cm,0cm}}
    \subfloat[]{
        \includegraphics[draft=false, width=0.23\textwidth, trim=2.5mm 3.5mm 2.5mm 2.9mm,clip]{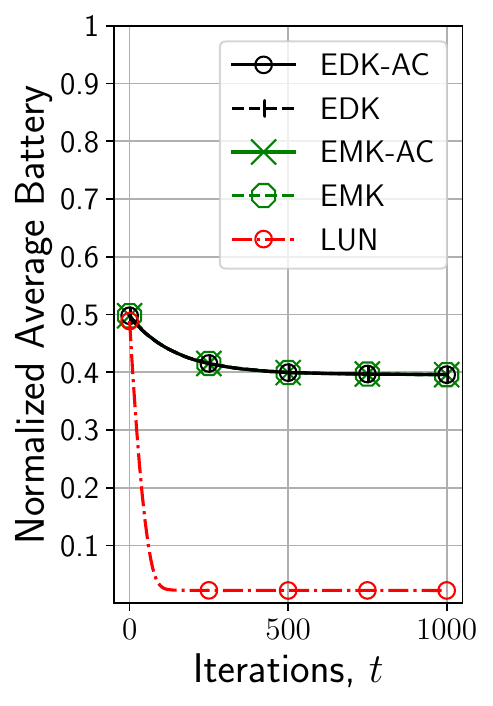}
        \label{fig:battery_vs_iterations}%
        }%
   \captionsetup[subfigure]{oneside,margin={1cm,0cm}}
    \subfloat[]{
        \includegraphics[width=0.236\textwidth, trim=2.5mm 3.55mm 2.7mm 2.5mm,clip]{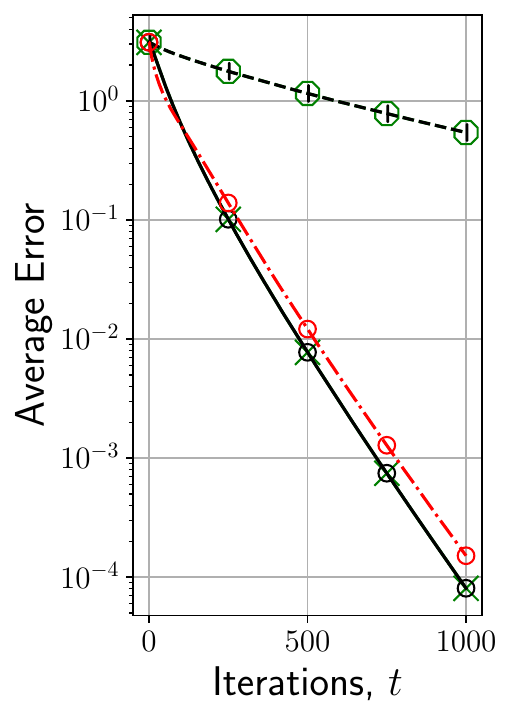}%
        \label{fig:error_vs_iterations}%
        }%
    \caption{(a) Normalized average battery level and (b) average error, i.e., \mbox{$\sum_k||\mathbf{w}_k(t) - \mathbf{w}||/K$}, as a function of the number of iterations for \mbox{$\xi = 0.4B_{\text{max}}$}, $m=0.2$, and $K=100$.}
\end{figure}

First, we set the desired threshold to $\xi = 0.4B_{\text{max}}$ and analyze the average stored energy over iterations in Fig.~\ref{fig:battery_vs_iterations}, which converges to the threshold when we exploit full knowledge of the energy income distribution (EDK; EDK-AC) or just its mean (EMK; EMK-AC). For the LUN approach, the average stored energy stabilizes near zero, as most users run out of energy. The network naturally reaches a stable state since all users, included those that run out of energy, continue to harvest energy. However, only users with sufficient energy actively participate in the training. Fig.~\ref{fig:error_vs_iterations} shows that relying solely on the energy income source, without  energy management, directly affects the learning process. Indeed, the LUN approach starts the training well, but soon devices die and are unable to resume learning until enough energy is harvested. Meanwhile, with the proposed energy management, devices can participate more frequently, resulting in a smaller error for EDK-AC and EMK-AC. Also, the error without the adaptive control is much higher, since it does not consider the norm of local updates, a persistent trend throughout the simulations.

\begin{figure}[t!] 
    \centering
    \captionsetup[subfigure]{oneside,margin={1cm,0cm}}
    \subfloat[]{%
        \includegraphics[width=0.39\textwidth,  trim=2.4mm 3.2mm 2.5mm 2.9mm,clip]{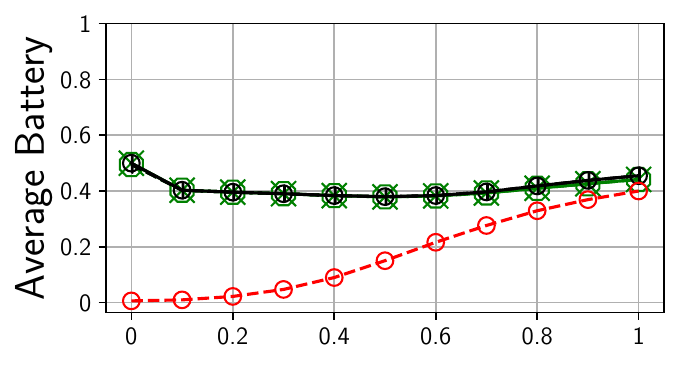}
        \label{fig:battery_vs_meanEnergyIncome}%
        }%
        \\
    \subfloat[]{%
        \includegraphics[width=0.41\textwidth, trim=2.4mm 2.6mm 2.3mm 2.4mm,clip]{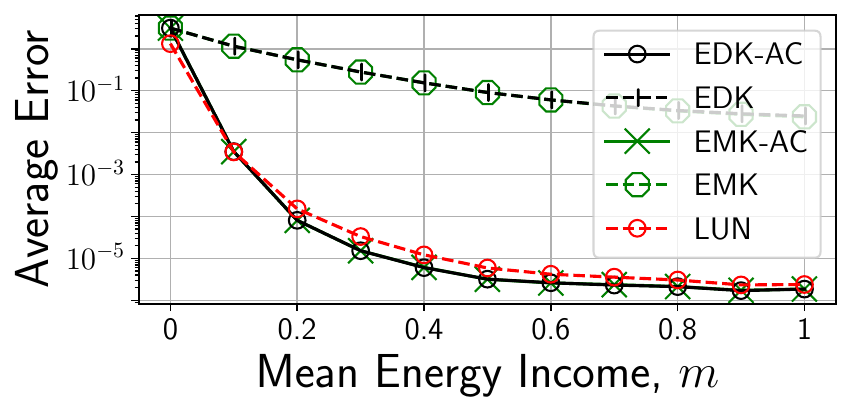}%
        \label{fig:error_vs_meanEnergyIncome}%
        }%
    \caption{(a) Normalized average battery level and (b) average error, i.e., \mbox{$\sum_k||\mathbf{w}_k(t) - \mathbf{w}||/K$}, versus the mean energy income for $\xi = 0.4B_{\text{max}}$, $t=1000$, and $K=100$.}
\end{figure}

Next we investigate the effect of the mean of the energy income process on the energy availability when $\xi = 0.4B_{\text{max}}$. Fig.~\ref{fig:battery_vs_meanEnergyIncome} displays the results for $t = 1000$, revealing that the EDK, EDK-AC, EMK, and EMK-AC curves stay fairly close to the threshold. The variation is due to the inequality in \eqref{eq:eh}, which, similar to the EMK approach, cannot fully incorporate the battery capacity considerations within this operational region. As we increase $m$, the EDK and EDK-AC curves depart from the EMK and EMK-AC curves, since the battery capacity limitation is more relevant. Besides, an energy surplus occurs within the network with respect to the threshold, since only $M$ devices transmit on average. In Fig.~\ref{fig:error_vs_meanEnergyIncome}, we plot the corresponding average error. For a small $m$, the threshold is too demanding, resulting in similar errors for all AC approaches. However, as the energy income increases, the proposed method with adaptive control outperforms  LUN. As the energy levels continue to rise, the differences between the AC methods and the LUN approach diminish.

\begin{figure}[!t] 
    \centering
    \captionsetup[subfigure]{oneside,margin={0.8cm,0cm}}
    \subfloat[]{
        \includegraphics[width=0.21\textwidth, trim=2.5mm 3.5mm 2.5mm 2.9mm,clip]{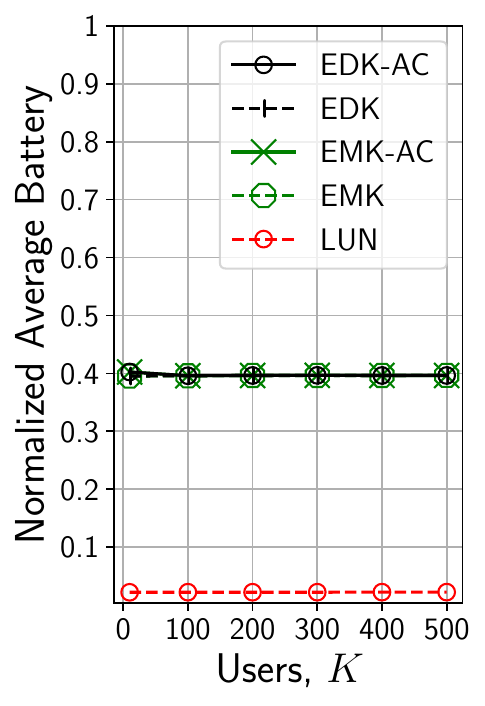}
        \label{fig:battery_vs_users}%
        }%
   \captionsetup[subfigure]{oneside,margin={1cm,0cm}}
    \subfloat[]{
        \includegraphics[width=0.216\textwidth, trim=2.5mm 3.5mm 2.6mm 2.5mm,clip]{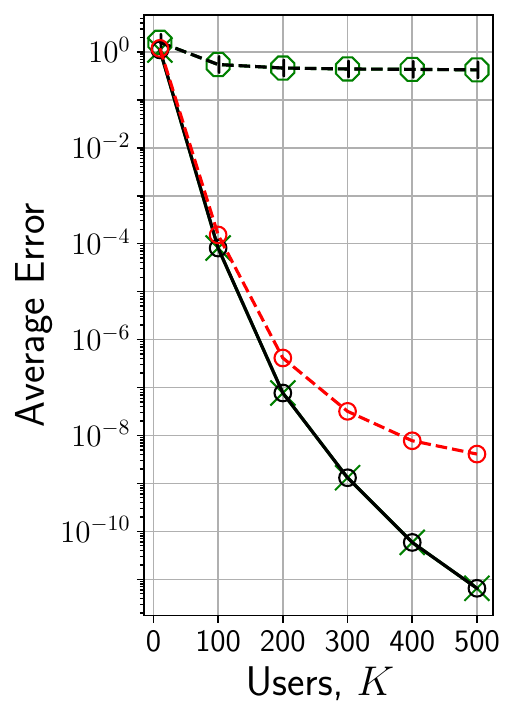}%
        \label{fig:error_vs_users}%
        }%
    \caption{(a) Normalized average battery level and (b) average error, i.e., \mbox{$\sum_k||\mathbf{w}_k(t) - \mathbf{w}||/K$}, as a function of the number of users for $\xi = 0.4B_{\text{max}}$, $m=0.2$, and $t=1000$.}
\end{figure}

In Fig.~\ref{fig:battery_vs_users} we set $m = 0.2$, $\xi=0.4B_{\text{max}}$, $t=1000$, for varying number of devices. The average battery level remains relatively unaffected, which is not true for the average error in Fig.~\ref{fig:error_vs_users}. Here, more users are able to engage in the learning process when using the proposed approaches. In contrast, the LUN method shows limited improvement with the number of users, since it lacks energy awareness, different from the methods that consider the average network energy. Thus, many users continue to consume energy by performing computations without transmitting, leading to rapid battery depletion. Moreover, since users in methods without AC have the same transmission probability, i.e., the methods disregard the informativeness of updates, the same performance improvements exhibited by methods with AC cannot be observed.

\begin{figure}[] 
    \centering
    \captionsetup[subfigure]{oneside,margin={0.8cm,0cm}}
    \subfloat[]{
        \includegraphics[width=0.21\textwidth, trim=2.5mm 3.3mm 2.3mm 2.5mm,clip]{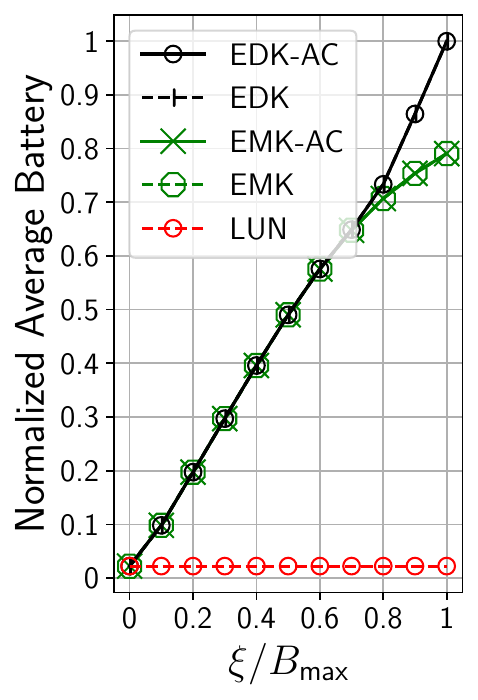}
        \label{fig:battery_vs_threshold}%
        }%
   \captionsetup[subfigure]{oneside,margin={1cm,0cm}}
    \subfloat[]{
        \includegraphics[width=0.21\textwidth, trim=2.5mm 3.3mm 2.5mm 2.5mm,clip]{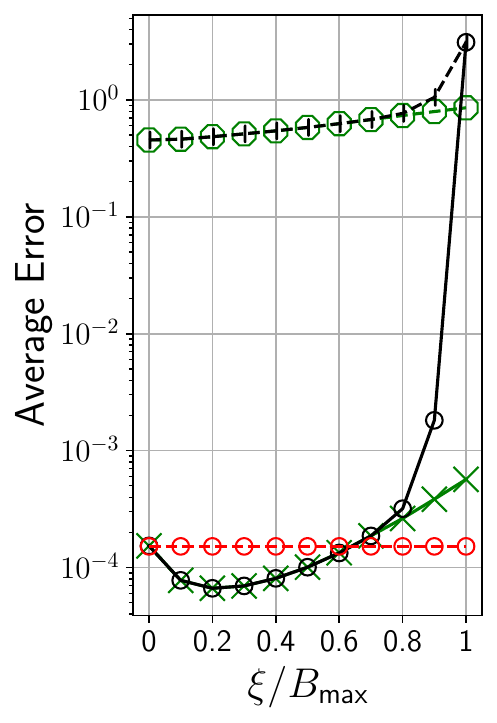}%
        \label{fig:error_vs_threshold}%
        }%
    \caption{(a) Normalized average battery level and (b) average error, i.e., \mbox{$\sum_k||\mathbf{w}_k(t) - \mathbf{w}||/K$}, as a function of the normalized threshold for $m=0.2$, $K=100$, and $t=1000$.}
\end{figure}

Finally, we examine the impact of the energy threshold. In Fig.~\ref{fig:battery_vs_threshold} it can be observed that the average battery level follows a nearly linear trend for EDK and EDK-AC, with slight variations due to \eqref{eq:eh}. When the threshold is set to lower or higher values, where the constraint is either insignificant or more dominant, the battery level precisely aligns with the threshold when using EDK and EDK-AC. However, with EMK and EMK-AC the battery cannot stabilize at the expected level for higher thresholds. As for the error, in Fig.~\ref{fig:error_vs_threshold}, it becomes apparent that an optimal threshold  exists, when considering the AC methods. If the threshold is too low, some devices deplete their energy and the error increases, while if the threshold is very demanding, the error rises since devices are often saving energy, reaching a point where LUN outperforms the proposed methods. It is worth mentioning that in the exceptional case where all users must maintain full battery, no training occurs as (energy-consuming) transmissions are not allowed. 

\section{Conclusion}\label{sec:conclusion}
We proposed an energy-aware method for FL networks under the principle of energy neutrality. Our approach mitigates battery depletion and achieves convergence to a sustainable energy level, enabling the execution of future tasks. The method requires distribution knowledge of the energy income, but relying only on average information was shown to be sufficient. In critical energy income regions and reasonable energy thresholds, our method outperforms the typical \mbox{norm-based} strategy, in terms of convergence time and battery level. In future works, we aim to include physical layer modeling and assess the impact of non-orthogonal multiple access techniques in the power domain and rate allocation procedures.

\bibliographystyle{IEEEtran}

\begin{thebibliography}{10}
\providecommand{\url}[1]{#1}
\csname url@samestyle\endcsname
\providecommand{\newblock}{\relax}
\providecommand{\bibinfo}[2]{#2}
\providecommand{\BIBentrySTDinterwordspacing}{\spaceskip=0pt\relax}
\providecommand{\BIBentryALTinterwordstretchfactor}{4}
\providecommand{\BIBentryALTinterwordspacing}{\spaceskip=\fontdimen2\font plus
\BIBentryALTinterwordstretchfactor\fontdimen3\font minus \fontdimen4\font\relax}
\providecommand{\BIBforeignlanguage}[2]{{%
\expandafter\ifx\csname l@#1\endcsname\relax
\typeout{** WARNING: IEEEtran.bst: No hyphenation pattern has been}%
\typeout{** loaded for the language `#1'. Using the pattern for}%
\typeout{** the default language instead.}%
\else
\language=\csname l@#1\endcsname
\fi
#2}}
\providecommand{\BIBdecl}{\relax}
\BIBdecl

\bibitem{Wahab2021}
O.~A. Wahab, A.~Mourad, H.~Otrok, and T.~Taleb, ``Federated machine learning: Survey, multi-level classification, desirable criteria and future directions in communication and networking systems,'' \emph{{IEEE} Commun. Surveys \& Tutorials}, vol.~23, no.~2, pp. 1342--1397, 2021.

\bibitem{Guo2021}
F.~Guo, F.~R. Yu, H.~Zhang, X.~Li, H.~Ji, and V.~C.~M. Leung, ``Enabling massive {I}o{T} toward 6{G}: A comprehensive survey,'' \emph{{IEEE} Internet of Things J.}, vol.~8, no.~15, pp. 11\,891--11\,915, 2021.

\bibitem{Albano2021}
M.~Albano, S.~Chessa, and K.~G. Larsen, ``A model-checking static analysis of task-based energy neutrality for energy harvesting {I}o{T},'' in \emph{IEEE Symp. Comp. Commu.}, 2021, pp. 1--7.

\bibitem{Choi2020}
J.~Choi and S.~R. Pokhrel, ``Federated learning with multichannel {ALOHA},'' \emph{{IEEE} Wireless Commun. Lett.}, vol.~9, no.~4, pp. 499--502, 2020.

\bibitem{daSilva2022}
R.~V. da~Silva, J.~Choi, J.~Park, G.~Brante, and R.~D. Souza, ``Multichannel {ALOHA} optimization for federated learning with multiple models,'' \emph{{IEEE} Wireless Commun. Lett.}, vol.~11, no.~10, pp. 2180--2184, 2022.

\bibitem{Wu2022}
H.~Wu and P.~Wang, ``Node selection toward faster convergence for federated learning on non-{IID} data,'' \emph{{IEEE} Trans. Netw. Sci. Eng.}, vol.~9, no.~5, pp. 3099--3111, 2022.

\bibitem{López2021}
O.~L.~A. López, H.~Alves, R.~D. Souza, S.~Montejo-Sánchez, E.~M.~G. Fernández, and M.~Latva-Aho, ``Massive wireless energy transfer: Enabling sustainable {I}o{T} toward 6{G} era,'' \emph{{IEEE} Internet of Things J.}, vol.~8, no.~11, pp. 8816--8835, 2021.

\bibitem{Zeng2022}
Q.~Zeng, Y.~Du, and K.~Huang, ``Wirelessly powered federated edge learning: Optimal tradeoffs between convergence and power transfer,'' \emph{{IEEE} Trans. Wireless Commun.}, vol.~21, no.~1, pp. 680--695, 2022.

\bibitem{Hamdi2022}
R.~Hamdi, M.~Chen, A.~B. Said, M.~Qaraqe, and H.~V. Poor, ``Federated learning over energy harvesting wireless networks,'' \emph{{IEEE} Internet of Things J.}, vol.~9, no.~1, pp. 92--103, 2022.

\bibitem{Konecny2017}
\BIBentryALTinterwordspacing
J.~Konecny, H.~B. McMahan, F.~X. Yu, P.~Richtarik, A.~T. Suresh, and D.~Bacon, ``Federated {Learning}: {Strategies} for {Improving} {Communication} {Efficiency},'' Oct. 2017, arXiv:1610.05492 [cs]. [Online]. Available: \url{http://arxiv.org/abs/1610.05492}
\BIBentrySTDinterwordspacing

\bibitem{Yao2023}
J.~Yao, Z.~Yang, W.~Xu, M.~Chen, and D.~Niyato, ``Gomore: Global model reuse for resource-constrained wireless federated learning,'' \emph{IEEE Wireless Communications Letters}, pp. 1--1, 2023.

\bibitem{Zhang2013}
W.~Zhang, Y.~Wen, K.~Guan, D.~Kilper, H.~Luo, and D.~O. Wu, ``Energy-optimal mobile cloud computing under stochastic wireless channel,'' \emph{{IEEE} Trans. Wireless Commun.}, vol.~12, no.~9, pp. 4569--4581, 2013.

\bibitem{Scaciota2022}
R.~Scaciota, O.~L.~A. López, G.~Brante, R.~D. Souza, A.~A. Mariano, and G.~L. Moritz, ``Batteryless machine-type communications with average channel state information energy beamforming,'' \emph{{IEEE} Acc.}, vol.~10, pp. 129\,676--129\,686, 2022.

\bibitem{Aygün2022}
O.~Aygün, M.~Kazemi, D.~Gündüz, and T.~M. Duman, ``Over-the-air federated learning with energy harvesting devices,'' in \emph{IEEE Glob. Commun. Conf.}, 2022, pp. 1942--1947.

\bibitem{Tamura2020}
M.~Tamura \emph{et~al.}, ``A 0.5 {V} {BLE} transceiver with a 1.9 {mW} {RX} achieving -96.4 {dBm} sensitivity and -27 {dBm} tolerance for intermodulation from interferers at 6 and 12 {MHz} offsets,'' \emph{{IEEE} J. Solid-State Circuits}, vol.~55, no.~12, pp. 3376--3386, 2020.

\bibitem{Xu2021}
Y.~Xu, H.~Zhou, J.~Chen, T.~Ma, and S.~Shen, ``Cybertwin assisted wireless asynchronous federated learning mechanism for edge computing,'' in \emph{IEEE Glob. Commun. Conf.}, 2021, pp. 1--6.

\bibitem{He2022}
J.~He, S.~Guo, D.~Qiao, and L.~Yi, ``Hete{FL}: Network-aware federated learning optimization in heterogeneous {MEC}-enabled {I}nternet of {T}hings,'' \emph{{IEEE} Internet of Things J.}, vol.~9, no.~15, pp. 14\,073--14\,086, 2022.

\end{thebibliography}
\mybibliography

\end{document}